\documentclass[aps,prd,nofootinbib,twocolumn,superscriptaddress,showpacs,floatfix]{revtex4}
\usepackage{mathrsfs}
\usepackage{latexsym}
\usepackage{amsmath}
\usepackage{amssymb}
\usepackage{graphicx}
\usepackage{longtable}

\usepackage{bbm}
\usepackage{epsfig}
\usepackage[usenames]{color}
\usepackage[colorlinks=true,citecolor=blue,linkcolor=blue]{hyperref}
\allowdisplaybreaks

\begin{document}



\title{ Baryon number fluctuations and QCD phase structure}

\author{Guo-yun Shao}
\email[Corresponding author: ]{gyshao@mail.xjtu.edu.cn}
\affiliation{School of Science, Xi'an Jiaotong
University, Xi'an, 710049, China}

\author{Zhan-duo Tang}
\affiliation{School of Science, Xi'an Jiaotong
University, Xi'an, 710049, China}


\author{Xue-yan Gao}
\affiliation{School of Science, Xi'an Jiaotong
University, Xi'an, 710049, China}

\author{Wei-bo He}
\affiliation{School of Science, Xi'an Jiaotong
University, Xi'an, 710049, China}


%







\begin{abstract}
We investigate the phase structure of strongly interacting matter and baryon number fluctuations in the Polyakov loop improved Nambu--Jona-Lasinio (PNJL)  model.  The calculation shows that both the chiral and deconfinement transitions, as well as their coincidence and separation determine the basic QCD phase structure. The contour maps and the three-dimensional diagrams of the net-baryon kurtosis and skewness present well the trace of QCD phase structure. Comparing with the experimental data, we find that the existence of a critical end point (CEP) of chiral transition is crucial to explain the non-monotonic energy dependence  and the large deviation from Poisson baseline of net-proton kurtosis. In particular, the relation between the chiral and deconfinement transitions in the crossover region is also reflected by the baryon number fluctuations. This study shows that the measurements of higher moments of multiplicity  distributions of conserved charges are powerful to investigate the criticality and even the chiral and deconfinement transitions in the crossover region. 
\end{abstract}


\maketitle

\section{Introduction}
The exploration of QCD phase diagram of strongly interacting matter
and the phase transition signatures are subjects of great interest in high energy nuclear physics. Intensive
searches on relativistic heavy-ion collision (HIC) have been performed in laboratories such as RHIC and LHC, and a near prefect fluid of quark-gluon plasma (QGP) has been created~\cite{Gupta11}. 
The properties of QGP will  be finally understood in the frame of quantum chromodynamics. 
However, in spite of tremendous theoretical and experimental
efforts, the QCD phase structure has not been unveiled yet due to its non-perturbation properties~\cite{Braun09,Fukushima11}.

Lattice QCD simulation is a fundamental tool to investigate the
thermodynamics of QCD matter at vanishing and small chemical
potential \cite{Karsch01, Karsch02,Kaczmarek05,Allton02,Cheng06,
Aoki09}, but it suffers the sign problem 
at large baryon chemical potential.
The region at large chemical potentials and low temperatures
essentially remains inaccessible although the great efforts that have been made\cite{Fodor07,Elia09,Ejiri08,Clark07}. 
On  the other hand, quantum field theories and  phenomenological  models,
such as the Dyson-Schwinger equation approach
\cite{Roberts00,Alkofer01,Cloet14,Fischer14, Gao16},
the Nambu--Jona-Lasinio (NJL) model \cite{Buballa05, Rehberg96,Alford08,Menezes14, Huang16}, the PNJL-type model
\cite{Fukushima04,Ratti06,Costa10,Fu08, Sasaki12, Ferreira14},
the Polyakov-loop extended quark-meson  (PQM) model
\cite{Schaefer10, Skokov11,Chatterjee12}, the Two-Phase model \cite {Shao11-2, Shao2015, Shao2016}
have been developed to give a complete description of QCD phase diagram.
However, a variety of  QCD phase structures are derived in different theories and models. A difficulty
is how to identify the genuine one from numerous candidates.

The most important observables for a thermal medium are fluctuations, which are powerful and unique to investigate the thermal properties of the medium. In HIC experiments, the fluctuations of net baryon number, electric charge and strangeness are most important observables to study the thermal properties of strongly interacting matter \cite{Aggarwal10, Adamczyk14, Abelev13, Adamczyk142}. The event-by-event measurements of statistical distributions of fluctuations of conserved charges have a primary mission in searching for QCD phase structure, in particular, pining down the existence the critical end point. Recently, in the first phase of  Beam Energy Scan (BES-I) program at RHIC a non-monotonic energy dependence of  the net-proton  kurtosis  $\kappa \sigma^2$ was observed, which shows a large deviation from the Poisson baseline and the prediction of the hadron resonance gas model \cite{Luo2014, Luo2016, Luo2017}. These behaviors possibly hint that the experiments  in Au+Au collisions at $\sqrt {s_{NN}}=7.7\,$GeV pass through the QCD critical region. To confirm the non-monotonic energy dependence of net-proton kurtosis, the second phase of RHIC beam energy scan (BES-II) is scheduled to take place during the years of 2019 and 2020 with more statistics data \cite{Luo2017}.There are also relevant experimental plans at NICA/FAIR/J-PARC in searching for the critical point and the QCD phase boundary.  The progress in experiments will provide a good opportunity to explore the properties of strongly interacting matter.

The  state of QGP created in HIC experiments  and the final observables at chemical freeze-out depend on the collision energies. In theory, when QGP matter is passing through the vicinity of the CEP, the fluctuations of conserved charges are severe and the high-order fluctuations can even change the sign,  oscillating on the two sides of the phase separation line \cite{Stephanov09, Fu10, Stephanov11, Schaefer12}.  That is why the measured large deviation of net-proton kurtosis from the baseline at RHIC is crucial to explore the QCD phase structure.  
Some theoretical and model calculations of the fluctuations of conserved quantities have been performed to explain the non-monotonic behavior of net-proton kurtosis \cite{ Friman11, Friman14, Swagato15, Chen17, Redlich17, Huang17, Fan16}. All these researches show that a CEP of chiral transition responsible for the large deviation of net-proton $ \kappa \sigma^2$ with the decreasing collision energies. 

In this study, we investigate the baryon number fluctuations up to the fourth order in the PNJL model, and study their relations with the chiral and deconfinement transitions.
The calculation shows both the chiral and deconfinement transitions are crucial to the baryon number fluctuations  not only in the critical region but also in the crossover region. The coincidence and separation of the chiral and deconfinement transitions determine the basic structure of QCD. By analyzing the fluctuation distributions of net baryon number, we find the trace of QCD phase structure closely related to the chiral and deconfinement phase transitions. We also display the three-dimensional  diagrams of kurtosis and skewness of net-baryon fluctuations. These diagrammatic presentations are very helpful to understand the relations between fluctuations and QCD phase structure.  


The paper is organized as follows. In Sec.~II, we introduce
briefly the thermodynamical description of fluctuations of conserved charges in a thermal system and the (2+1) flavor PNJL quark model. In Sec.~III, we present the numerical results of the QCD phase diagram and the statistic distributions of fluctuations of conserved quantities. We then discuss the relation between fluctuations of conserved charges and QCD phase structure. A summary is finally given in Sec. IV.

\section{ Fluctuations of conserved charges}
We first give a brief introduction of thermal fluctuations of conserved charges which are important observables in experiments in searching for the QCD phase structure. For a grand-canonical ensemble, the pressure of the thermal system is related to the logarithm of the partition function \cite{Karsch15}:
\begin{equation}\label{}
\frac{P}{T^4}=\frac{1}{VT^3}\ln[Z(V,T,\mu_B, \mu_Q, \mu_S)],
\end{equation}  
where $V$ and $T$ are the volume and temperature of the system. The $\mu_B, \mu_Q, \mu_S$ are the chemical potentials of baryon number, electric charge and strangeness, respectively. The generalized susceptibilities of conserved charges can be derived by taking the partial derivatives of the pressure with respect to the corresponding chemical potentials \cite{Luo2017}
\begin{equation}\label{}
\chi^{BQS}_{ijk}=\frac{\partial^{i+j+k}[P/T^4]}{\partial(\mu_B/T)^i \partial(\mu_Q/T)^j \partial(\mu_S/T)^k},
\end{equation}  

The cumulants of multiplicity distributions of the conserved charges are connected with the generalized susceptibilities by
 \begin{equation}\label{}
\!C^{BQS}_{ijk}\!=\!\frac{\partial^{i+j+k}\ln[Z(V,T,\mu_B, \mu_Q, \mu_S)]}{\partial(\mu_B/T)^i \partial(\mu_Q/T)^j \partial(\mu_S/T)^k}\!=\!VT^3\chi^{BQS}_{ijk}\,
\end{equation} 
In experiments, observables are constructed by the ratio of cumulants, which cancel the volume dependence and can be compared with theoretical calculations of the generalized susceptibilities.

For an arbitrary distribution we can define the Gaussian width $\sigma^2$ with the non-Gaussian fluctuations. In this study, we focus on two important statistic quantities, the skewness ($S\sigma$) and Kurtosis ($\kappa \sigma^2$). For the net baryon number probability distributions, the skewness and Kurtosis are related to the high order cumulants as
\begin{equation}\label{}
S\sigma=\frac{C^B_3}{C^B_2}=\frac{\chi^B_3}{\chi^B_2} \,\,\,\,\,\texttt{and}\,\,\,\,  \kappa \sigma^2=\frac{C^B_4}{C^B_2}=\frac{\chi^B_4}{\chi^B_2}.
\end{equation}  
Similar relations can be derived for the fluctuations of electric charge and strangeness. 

We will calculate the susceptibilities of conserved charges in the (2+1) flavor PNJL quark model. The chemical potentials $\mu_B, \mu_Q, \mu_S$ used in experiments and Lattice QCD simulations are related to the quark chemical potentials with the following relations
\begin{equation}\label{chemical_1}
\mu_u=\frac{1}{3}\mu_B+\frac{2}{3}\mu_Q, \,\,\,\,\,\,\,\,\,\,\mu_d=\frac{1}{3}\mu_B-\frac{1}{3}\mu_Q,
\end{equation}  
and 
\begin{equation}\label{chemical_2}
\mu_s=\frac{1}{3}\mu_B-\frac{1}{3}\mu_Q-\mu_S,
\end{equation}  
where $\mu_{u,d,s}$ are the quark chemical potentials for up, down and strange quarks.

The Lagrangian density in the three-flavor PNJL model is taken as
\begin{eqnarray}
\mathcal{L}&\!=&\!\bar{q}(i\gamma^{\mu}D_{\mu}\!+\!\gamma_0\hat{\mu}\!-\!\hat{m}_{0})q\!+\!
G\sum_{k=0}^{8}\big[(\bar{q}\lambda_{k}q)^{2}\!+\!
(\bar{q}i\gamma_{5}\lambda_{k}q)^{2}\big]\nonumber \\
           &&-K\big[\texttt{det}_{f}(\bar{q}(1+\gamma_{5})q)+\texttt{det}_{f}
(\bar{q}(1-\gamma_{5})q)\big]\nonumber \\ \nonumber \\
&&-\mathcal{U}(\Phi[A],\bar{\Phi}[A],T),
\end{eqnarray}
where $q$ denotes the quark fields with three flavors, $u,\ d$, and
$s$; $\hat{m}_{0}=\texttt{diag}(m_{u},\ m_{d},\
m_{s})$ in flavor space; $G$ and $K$ are the four-point and
six-point interacting constants, respectively.  The $\hat{\mu}=diag(\mu_u,\mu_d,\mu_s)$ are the quark chemical potentials which are related to chemical potentials of the conserved charges through Eq.~(\ref{chemical_1}) and (\ref{chemical_2}).

The covariant derivative in the Lagrangian is defined as $D_\mu=\partial_\mu-iA_\mu$.
The gluon background field $A_\mu=\delta_\mu^0A_0$ is supposed to be homogeneous
and static, with  $A_0=g\mathcal{A}_0^\alpha \frac{\lambda^\alpha}{2}$, where
$\frac{\lambda^\alpha}{2}$ is $SU(3)$ color generators.
The effective potential $\mathcal{U}(\Phi[A],\bar{\Phi}[A],T)$ is expressed in terms of the traced Polyakov loop
$\Phi=(\mathrm{Tr}_c L)/N_C$ and its conjugate
$\bar{\Phi}=(\mathrm{Tr}_c L^\dag)/N_C$. The Polyakov loop $L$  is a matrix in color space
\begin{equation}
   L(\vec{x})=\mathcal{P} exp\bigg[i\int_0^\beta d\tau A_4 (\vec{x},\tau)   \bigg],
\end{equation}
where $\beta=1/T$ is the inverse of temperature and $A_4=iA_0$.

The Polyakov-loop effective potential used in this study is given in ~\cite{Robner07},
%
\begin{eqnarray}
     \frac{\mathcal{U}(\Phi,\bar{\Phi},T)}{T^4}&=&-\frac{a(T)}{2}\bar{\Phi}\Phi +b(T)\mathrm{ln}\big[1-6\bar{\Phi}\Phi\\ \nonumber
                                                &&+4(\bar{\Phi}^3+\Phi^3)-3(\bar{\Phi}\Phi)^2\big],
\end{eqnarray}
where
\begin{equation}
   \!a(T)\!=\!a_0\!+\!a_1\big(\frac{T_0}{T}\big)\!+\!a_2\big(\frac{T_0}{T}\big)^2 \,\,\,\texttt{and}\,\,\,\,\, b(T)\!=\!b_3\big(\frac{T_0}{T}\big)^3.
\end{equation}
The parameters $a_i$, $b_i$ listed in Table. \ref{tab:1} are fitted according to the lattice simulation of  QCD thermodynamics in
pure gauge sector. And $T_0$ is found to be 270 MeV as the critical temperature for the deconfinement phase transition of gluon part at zero chemical potential~\cite{Fukugita90}. When fermion fields are included, a rescaling of $T_0=210$\, Mev 
is implemented to obtain a consistent result between model calculation and full lattice simulation.
\begin{table}[ht]
\tabcolsep 0pt \caption{\label{tab:1}Parameters in the Polyakov-loop potential~\cite{Robner07}}
\setlength{\tabcolsep}{21pt}
\begin{center}
\def\temptablewidth{0.8\textwidth}
\begin{tabular}{c c c c}
\hline
\hline
   {$a_0$}                      & $a_1$        & $a_2$      & $a_3$           \\  \hline
   $ 3.51$                   & -2.47        &  15.2      & -1.75               \\ \hline
\hline
\end{tabular}
\end{center}
\end{table}

In the mean field approximation, the constituent quark masses $M_i$
are obtained as
\begin{equation}
M_{i}=m_{i}-4G\phi_i+2K\phi_j\phi_k\ \ \ \ \ \ (i\neq j\neq k),
\label{mass}
\end{equation}
where $\phi_i$ stands for quark condensate of the flavor $i$.
The thermodynamical potential  of the PNJL model is derived as
\begin{widetext}
\begin{eqnarray}
\Omega&=&\mathcal{U}(\bar{\Phi}, \Phi, T)+2G\left({\phi_{u}}^{2}
+{\phi_{d}}^{2}+{\phi_{s}}^{2}\right)-4K\phi_{u}\,\phi_{d}\,\phi_{s}-2\int_\Lambda \frac{\mathrm{d}^{3}p}{(2\pi)^{3}}3(E_u+E_d+E_s) -2T \sum_{u,d,s}\int \frac{\mathrm{d}^{3}p}{(2\pi)^{3}} \big[\mathrm{ln}(1\nonumber \\
&&+3\Phi e^{-(E_i-\mu_i)/T}+3\bar{\Phi} e^{-2(E_i-\mu_i)/T}+e^{-3(E_i-\mu_i)/T}) \big]
-2T \sum_{u,d,s}\int \frac{\mathrm{d}^{3}p}{(2\pi)^{3}} \big[\mathrm{ln}(1+3\bar{\Phi} e^{-(E_i+\mu_i)/T}\nonumber \\
&&+3\Phi e^{-2(E_i+\mu_i)/T}+e^{-3(E_i+\mu_i)/T}) \big],
\end{eqnarray}
\end{widetext}
where $E_i=\sqrt{\vec{p}^{\,2}+M_i^2}$ is the energy-momentum dispersion relation.

The values of $\phi_u, \phi_d, \phi_s, \Phi$ and $\bar{\Phi}$ are determined by minimizing the thermodynamical
potential
\begin{equation}
\frac{\partial\Omega}{\partial\phi_u}=\frac{\partial\Omega}{\partial\phi_d}=\frac{\partial\Omega}{\partial\phi_s}=\frac{\partial\Omega}{\partial\Phi}=\frac{\partial\Omega}{\partial\bar\Phi}=0.
\end{equation}
All the thermodynamic quantities relevant to the bulk properties of quark matter can be obtained from $\Omega$. Especially, the pressure
and energy density should be zero in the vacuum.

In the calculation a cut-off $\Lambda$ is implemented in 3-momentum
space for divergent integrations. We take the model parameters obtained in ~Ref.~\cite{Rehberg96}:
$\Lambda=603.2$ MeV, $G\Lambda^{2}=1.835$, $K\Lambda^{5}=12.36$,
$m_{u,d}=5.5$  and $m_{s}=140.7$ MeV, determined
by fitting $f_{\pi}=92.4$ MeV,  $M_{\pi}=135.0$ MeV, $m_{K}=497.7$ MeV and $m_{\eta}=957.8$ MeV of their
experimental values.

\section{Numerical results and discussions }
In this section,  we first analyze the
QCD phase structure in the improved PNJL model. We then investigate the statistical distributions of baryon number fluctuations and discuss their relations with the phase transitions including both the chiral and deconfinement  transitions. In the calculation, we take $\mu_Q=\mu_S=0$ for
simplicity, which is also approximately consistent with the experimental measurements~\cite{Cleymans05}. Strange quark chemical potential is set to zero due to the conservation of strangeness in the process of strong interaction.  

\subsection{QCD phase structure in the  PNJL model}
There have been already  numerous researches on the QCD phase structure. Here we  just focus on some respects crucial to the observables of fluctuations of conserved charges.

\begin{figure}[htbp]
\begin{center}
\includegraphics[scale=0.3]{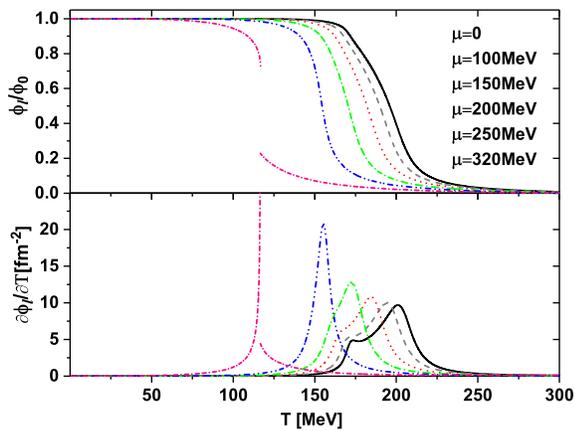}
\caption{\label{fig:chiral-condensate}(color online) Quark condensate in the upper panel, the partial derivative of $\phi_l$ respect to temperature in the lower panel. }
\end{center}
\end{figure}

We show  the chiral phase transition along the temperature for various quark chemical potentials in Fig.~\ref{fig:chiral-condensate}. The lines in the the upper panel present the variation of quark condensate $\phi_l$ ($\phi_l=\phi_u=\phi_d $ for symmetric quark matter) as functions of temperature for given quark chemical potentials. It shows that $\phi_l/\phi_0$ decreases continuously when quark chemical potential $\mu$ is equal to $0, 100, 200$ and $250\,$MeV, which means the chiral phase transition is a smooth crossover.  For $\mu=320\,$MeV, there is a jump in the curve of chiral condensate, indicating the occurring of the first-order transition.  The feature of the chiral first-order transition has been detailly discussed in the literature (e.g. \cite{Fukushima04,Ratti06,Costa10}). Here we pay more attention to the chiral crossover and deconfinement transitions which have a close relation to the fluctuations of conserved charges at small and intermediate chemical potentials.

The lower panel of Fig.~\ref{fig:chiral-condensate} shows the partial derivatives of $\phi_l$ respect to temperature  which indicates how fast the chiral phase transition is with the increase of temperature for given quark chemical potentials. One evident features  for $\mu=0$ is that  $\partial{\phi_l}/\partial T$ has two maxima. This structure only exists at small chemical potentials. The smaller maximum on the lower-temperature side will gradually disappear with the increase of quark chemical potential, as shown in the curves of $\mu_q=100$ and $150\,$MeV, but the widths of the transition on both sides of the stronger peak are asymmetric. Finally, a single peak with almost symmetric transition width on both sides forms, as presented by the curve of $\mu=250\,$MeV. If the peaks of the larger maxima in the $T-\mu$ plane are connected,  the chiral crossover separation line can be derived, which will be plotted later in the QCD phase diagram.  

\begin{figure}[htbp]
\begin{center}
\includegraphics[scale=0.38]{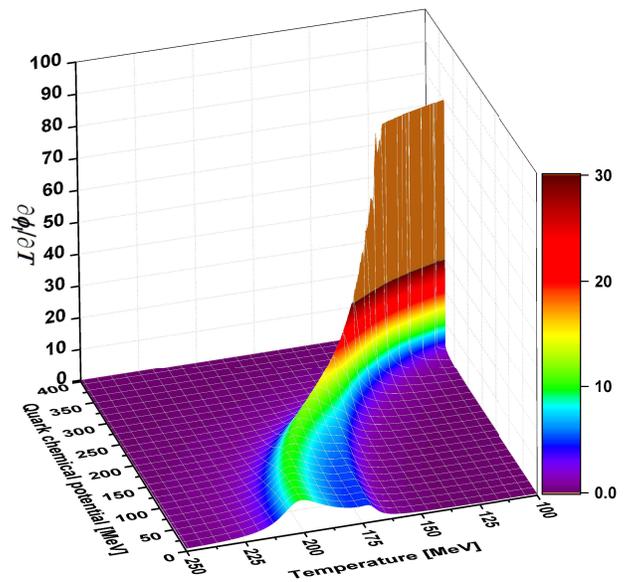}
\caption{\label{fig:dphi-dT}(color online) Three-dimensional structure of $\partial{\phi_l}/\partial T$.}
\end{center}
\end{figure}

For the convenience of later discussion of baryon number fluctuations, we visualize the phase structure of chiral transition in the three-dimensional diagram  in Fig.~\ref{fig:dphi-dT}. This figure clearly demonstrates that how fast the chiral crossover  transition takes place as a function of temperature and  chemical potential. It also indicates the appearance  of the first-order phase transition at larger chemical potentials where $\partial{\phi_l}/\partial T$ diverges.

\begin{figure}[htbp]
\begin{center}
\includegraphics[scale=0.3]{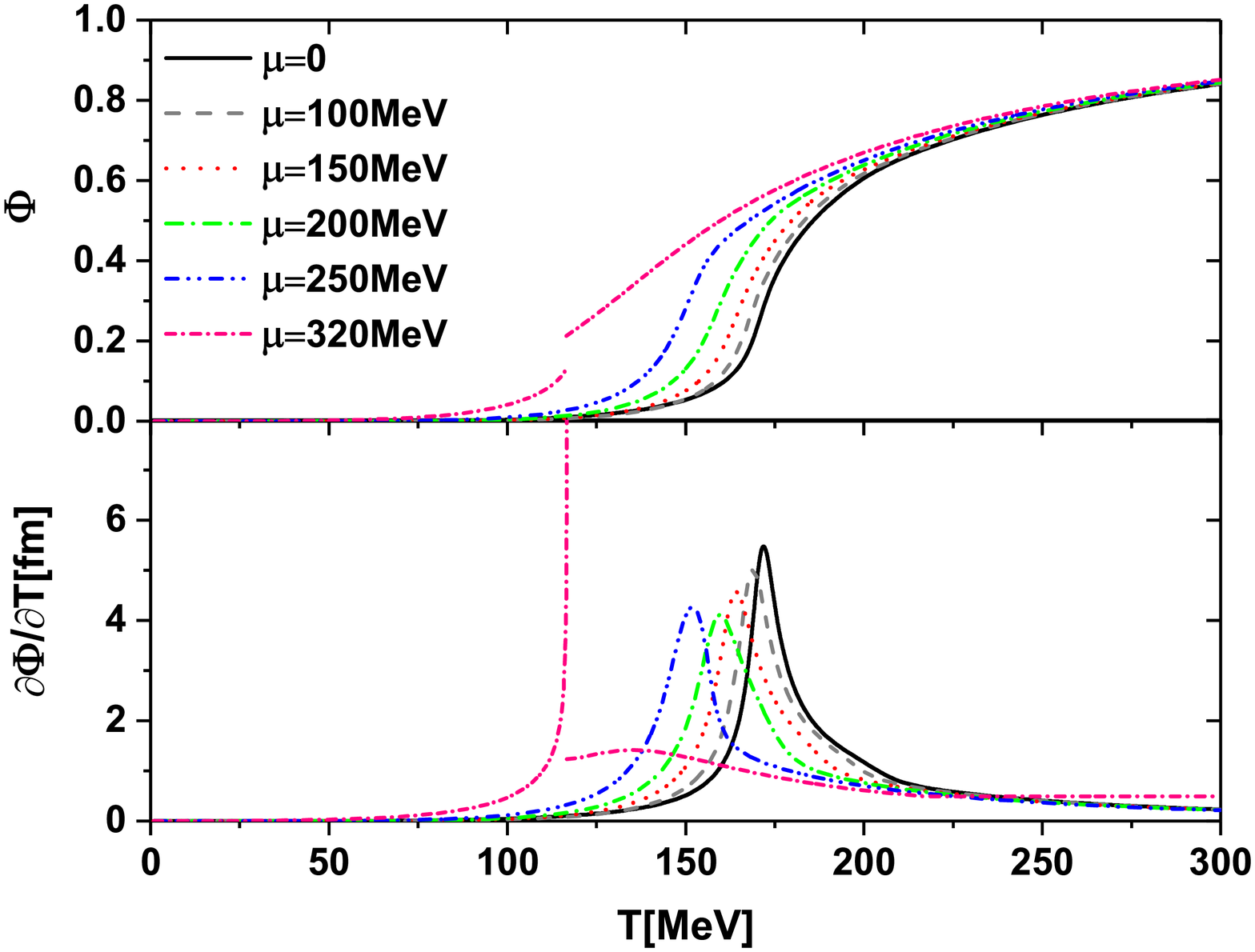}
\caption{\label{fig:deconfinement.eps}(color online) Polyakov loop $\Phi$ (upper panel) and its partial derivative respect to temperature $\partial{\Phi}/\partial T$ (lower panel). }
\end{center}
\end{figure}

$\Phi$ and $\bar{\Phi}$ are the quantities to describe the cofinement-deconfinement phase transition.  The values of them are equal to each other at vanishing chemical potential. Their difference are quite small even at finite chemical potential. Therefore, here we just plot in Fig.~\ref{fig:deconfinement.eps} the value of $\Phi$ and its derivative respect to temperature.  The upper (lower) panel shows that  $\Phi$ ($\partial{\Phi}/\partial T$) changes continuously for $\mu=0, 100, 150, 200$ and $250\,$MeV.  There is a jump of $\Phi$ for $\mu=320\,$MeV, which is induced by the restoration of chiral symmetry with a first-order transition. 

The confinement-deconfinement phase transition line is usually derived with the requirement of ($\Phi+\bar{\Phi})/2=0.5$ in literature. Alternatively, we can obtain a phase transition line by the requirment that $\partial{\Phi}/\partial T$ takes the maximum for a given $\mu$. In this method, there is a special case that we need to pay attention to. There is a jump of $\Phi$ along the first-order phase transition line.  Although $\partial{\Phi}/\partial T$ diverges in this case for $\mu$ much larger than $\mu_{_C}$,  the deconfinement still does not occur. The main reason is that the value of $\Phi$ is still quite small along the first-order transition. As a matter of fact, there also exists the other local maximum of $\partial{\Phi}/\partial T$ at a temperature higher than the jump point of  $\Phi$ as shown in Fig.~\ref{fig:dphi-dT}. If we connect these points with those obtained in the crossover region by the requirement of $\partial{\Phi}/\partial T$ taking the local maximum, we can obtain the deconfinement transition line in the full QCD phase diagram. 

\begin{figure}[htbp]
\begin{center}
\includegraphics[scale=0.38]{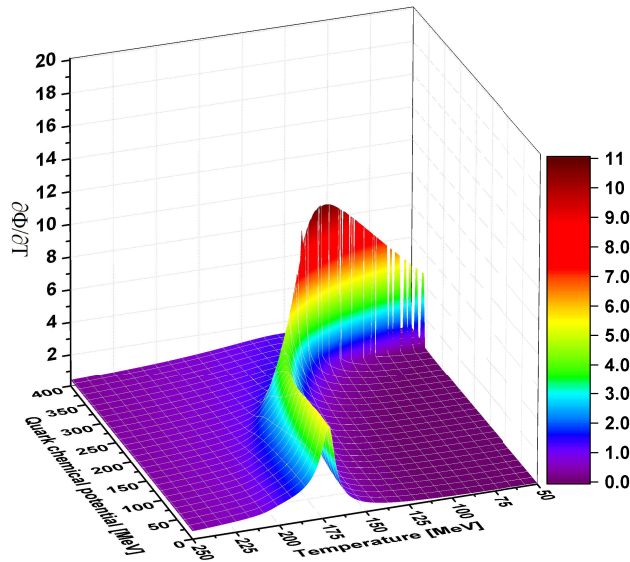}
\caption{\label{fig:dPHIdT}(color online) Three-dimensional structure of $\partial{\Phi}/\partial T$.}
\end{center}
\end{figure}

The partial derivative of  $\Phi$ respect to $T$, shown in the lower panel of Fig.~\ref{fig:deconfinement.eps}, reflects the width of the deconfinement transition. 
We visualize $\partial{\Phi}/\partial T$ as a function of temperature and chemical potential  in a three-dimensional diagram in Fig.~\ref{fig:dPHIdT}, which
clearly presents the features of the deconfinement transition. We notice that the peak of $\partial{\Phi}/\partial T$ becomes quite flat at higher temperatures above the first-order transition line for $\mu>\mu_{_{C}}$.

\begin{figure}[htbp]
\begin{center}
\includegraphics[scale=0.32]{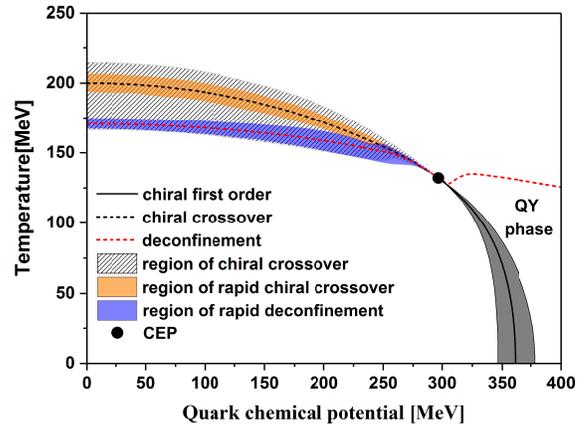}
\caption{\label{fig:full-qcd-diagram}(color online) QCD phase diagram. The black solid line is the chiral first-order transition line, and the black dash line is that of chiral crossover. The red dash curve is the deconfinement transition line obtained with $\partial{\Phi}/\partial T$ taking the maximum. The area filled with oblique lines is roughly the region of chiral crossover, and the transition is more rapid in the yellow band (approximately corresponding to the green peak in Fig.~\ref{fig:dphi-dT}) than the rest part. The blue band is the rapid deconfinement transition region. ``QY" stands for quarkyonic phase. }
\end{center}
\end{figure}

Fig.~\ref{fig:chiral-condensate}\,--\,\ref{fig:dPHIdT} indicate that the chiral and deconfinement  transitions at small
chemical potentials occur in wide a range of temperature and they overlap in some areas.  
To indicate the relations between the chiral and deconfinement transitions we plot the complete phase diagram in Fig.~\ref{fig:full-qcd-diagram}. 
The black solid line is the chiral first-order transition line, and the black dash line is that of the chiral crossover transition. The red dash curve is the deconfinement transition line obtained with $\partial{\Phi}/\partial T$ taking the maximum for a given chemical potential. The area filled with oblique lines is roughly the region of chiral crossover, and the transition is more rapid in the yellow band (corresponding to the green peak in Fig.~\ref{fig:dphi-dT}) than in the rest part. However, for the deconfinement transition, the rapid transition region lies in the blue band.  Comparing with the chiral transition, we find the blue band of the rapid deconfinement is almost overlap with the smaller peak of chiral transition, as shown in Fig.~\ref{fig:chiral-condensate} and Fig.~\ref{fig:dphi-dT}.

Fig.~\ref{fig:full-qcd-diagram}   shows that the chiral and deconfinement transitions can be approximately separated at  small chemical potentials, but they almost overlap near the critical region. According to the features above, we can come to a conclusion that the partial restoration of chiral symmetry near the smaller peak is induced to a certain degree by the deconfinement transition, but the deconfinement transition is not strong enough to force the complete restoration of chiral symmetry.  It becomes difficult to separate the two kinds of phase transition with the increase of chemical potential, in particular, in the critical region where the transition occurs first  will trigger the other.

Fig.~\ref{fig:full-qcd-diagram} also indicates that the Polyakov loop $\Phi$ is not very sensitive to the chemical potential with the taking place of the chiral first-order transition. The chiral and deconfinment transitions separate again. This feature leads to the so-called quarkyonic (QY) phase in which the chiral symmetry is restored but quarks are still confined. The area of  quarkyonic phase will shrink when the quantum back-reaction of
the matter sector to the gluonic sector is considered, which is not the point we emphasize in this study. For the spinodal structure (the gray area in Fig.~\ref{fig:full-qcd-diagram}) of chiral first-order transition, one can refer to \cite{Costa10} for a detailed description.

\section{Baryon number fluctuations}
The fluctuations of conserved  charges are most important observables to investigate the thermal properties of strongly interacting matter. The higher order moments  of various particles multiplicities distributions can be measured event-by-event. The experimental data of  cumulants  can serve as a probe to explore the QCD phase structure, in particular the characteristic signatures of critical behavior.  In this subsection,  we present the numerical results of net-baryon number fluctuation distributions and compare them with the experimental data.

\begin{figure}[htbp]
\begin{center}
\includegraphics[scale=0.35]{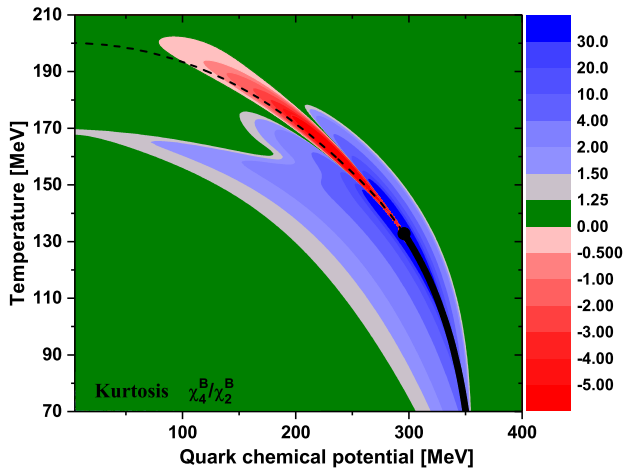}
\caption{\label{fig:kurtosis-2D-B}(color online) Contour lines of the  net-baryon kurtosis in the $T-\mu$
 plane. $\kappa \sigma^2$  is negative in the red area and positive in the rest area.}
\end{center}
\end{figure}

\begin{figure}[htbp]
\begin{center}
\includegraphics[scale=0.35]{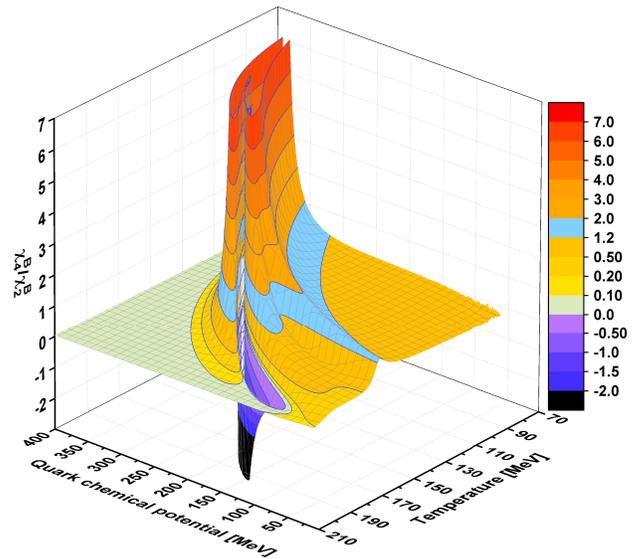}
\caption{\label{fig:kurtosis-3D-B}(color online) Three-dimensional structure of  net-baryon kurtosis as a function of temperature and chemical potential.}
\end{center}
\end{figure}

We  show  the kurtosis  $\kappa \sigma^2= \chi_{4}^B/\chi_{2}^B$ of net-baryon fluctuation distributions in Fig.~\ref{fig:kurtosis-2D-B} and Fig.~\ref{fig:kurtosis-3D-B}. 
The two figures present the contour map and three-dimensional landscape of  kurtosis, repectively.
 Fig.~\ref{fig:kurtosis-2D-B} exhibits that there are two branches of the contour map of the net-baryon number kurtosis, and they converge near the critical region. One branch is along the chiral transition line, as derived in the NJL model\cite{Chen17, Fan16} (Only one branch exists in the NJL model). 
For this branch, the value of $\kappa \sigma^2$ is universally negative in a narrow region (red area in Fig.~\ref{fig:kurtosis-2D-B}) when the critical point is approached on the crossover side of the chiral transition line. The kurtosis oscillates more and more severely when the chiral crossover transition line is passed through from one side to the other near the critical region.  These characteristic behaviors are more impressively demonstrated in Fig.~\ref{fig:kurtosis-3D-B}. It is like a bottomless valley sandwiched between two steep mountains. 

The numerical results show that  $\kappa \sigma^2$ diverge near the critical region. However, the divergence of kurtosis will not appear in experiments even the chemical freeze-out is very close to the critical end point, because the observed particle number in one collision is finite.
Thanks to the extensive measurement of BES I program,  a significant deviation of the kurtosis of net-proton multiplicity distribution from the Poission baseline has been observed. The statistic central value of   $\kappa \sigma^2$  in experiments can reach about 3.5 in Au+Au collision with the collission energy $\sqrt{s_{_{NN}}}=7.7\,$GeV for the centrality 0-5\% \cite{Luo2014, Luo2016, Luo2017}. Considering the expansion after the phase transition, the fluctuation signatures measured is in fact weakened. 

Comparing with experimental data with the contour lines  and three-dimensional structure of $\kappa \sigma^2$, we can imagine that a region close to the CEP is reached in Au+Au collision with $\sqrt{s_{_{NN}}}=7.7\,$GeV.  Since in the Poisson distribution $\kappa \sigma^2$ is expected to be unity in a theory with the absence of criticality.
If the significant deviation of kurtosis of net-proton number from  the Poission baseline
can be confirmed by more statistical data in the  BES II program scheduled to take place during the years of 2019-2010 focusing on 7.7-19.6 GeV,  a decisive conclusion will be that the critical region has been approached to.  To investigate the behavior of $\kappa \sigma^2$ at even larger chemical potential,  experiments should be run with lower collision energies than $7.7\,$GeV. The fixed-target experiment at RHIC and  Compressed Baryonic Experiment (CMB) at FAIR will play important roles on this respect \cite{Luo2017}.

\begin{figure}[htbp]
\begin{center}
\includegraphics[scale=0.35]{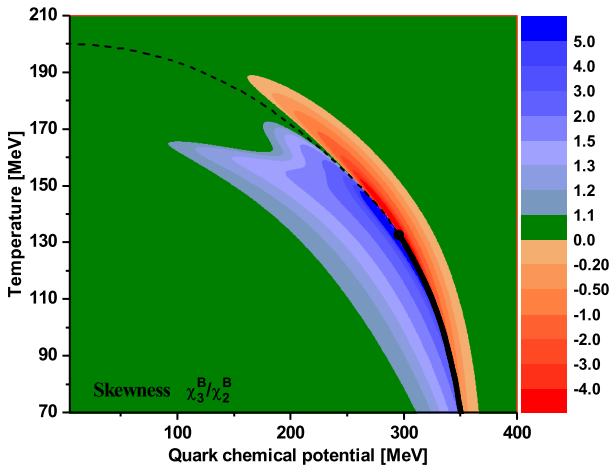}
\caption{\label{fig:skewness-2D-B}(color online)  Contour lines of skewness of net-baryon number fluctuation in the $T-\mu$
 plane. $S \sigma$  is negative in the red region and positive in the rest region.}
\end{center}
\end{figure}

The second branch of $\kappa \sigma^2$ of net-baryon number fluctuation  appears only in the crossover region. The peaks of the contour lines in this branch lie in the region of rapid deconfinement transition, as shown in figure~5.  Such a structure  also emerges in the three-dimensional diagram. 
The appearance of this branch can be attributed  to the  deconfinement  transition which also induces the partial restoration of chiral symmetry, as discussed in the last subsection.  
The existence of two branches of $\kappa \sigma^2$ indicates that both the chiral and deconfinement transitions are important to the  observables of fluctuations. This involves the coincidence and/or separation of the chiral and deconfinement transitions at vanishing and finite chemical potential, which is still an open issue \cite{Coleman80, Fischer142, Xin14}.  Although one can regulate the relation between the two phase transitions by introducing some new interactions or adjusting some model parameters, it is not the point we emphasize in this study.  The point is that the structure of strongly interacting matter related  to both the chiral and deconfinement transitions can be inferred from the fluctuations of conserved charges, which points out a direction to explore the QCD phase structure. Comprehensive researches in this aspect need to be carried out in both theories and experiments . 

\begin{figure}[htbp]
\begin{center}
\includegraphics[scale=0.35]{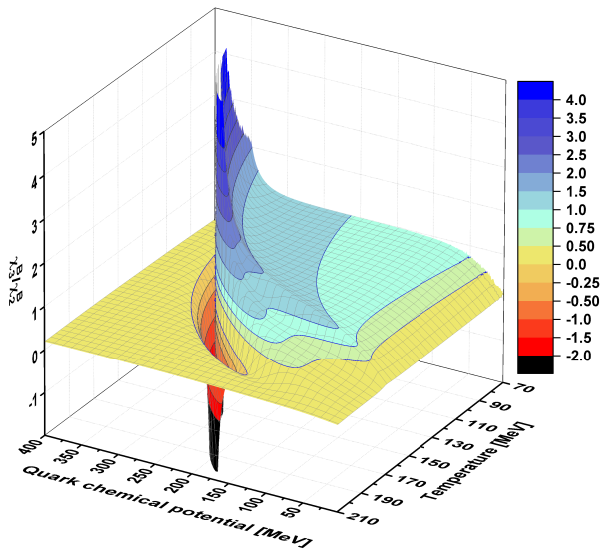}
\caption{\label{fig:skewness-3D-B}(color online) Three-dimensional structure of kurtosis of net-baryon number a fluctuation as function of temperature and chemical potential.}
\end{center}
\end{figure}

We  demonstrate the contour map of net-baryon skewness $S\sigma$ in Fig.~\ref{fig:skewness-2D-B}. It indicates that the value of  $S\sigma$ is negative  in a slender region close to the chiral separation line on the right side.  
The value of  $S\sigma$ also diverges near the critical region. Fig.~\ref{fig:skewness-3D-B} demonstrates the three-dimensional structure of the distribution of $S\sigma$ as a function of temperature and chemical potential.
It is  easy to understand the increase of  net-proton $S\sigma$  with the decrease of collision energy \cite{Luo2014, Luo2016, Luo2017}.
Similar to  the distribution of $\kappa \sigma^2$, there are also two branches existing.  In general, the value of  $S\sigma$  is smaller than $\kappa \sigma^2$, therefore the measurement of $\kappa \sigma^2$ is more effective to investigate the criticality.

Finally, we note that the net proton number is not a conserved charge. The comparison of baryon number fluctuations and net proton fluctuations measured in experiments is rough, which is feasible assuming that proton and neutron are independent as in the calculation of HRG model.  A dynamical origin of the isospin correlations has been studied in~\cite{Kitazawa12} assuming  the  complete randomization of isospin of nucleons in the final state in HIC experiments.  The residual interactions
after the chemical freeze-out can change $p$ into $n$ and vice versa. Based on the  formula derived in  
the data analysis shows that the $\kappa \sigma^2$ of net baryon number are systematically lower than that of net proton results~\cite{Luo2017}.
However, the research in~\cite{fukushima15} find that isospin correlation can produce about 10\% correction, not enough to explain the suppression tendency in the kurtosis at lower collision energies.  Besides, the correlation among proton number and neutron number is possibly  induced by the flavor mixing through the off-diagonal susceptibility $\chi_{ud}$ arising from $\pi$ dynamics.  A further study in  this direction is necessary to search for the relationship between baryon number fluctuations and proton number fluctuations.

\section{summary}
In this research, we investigate the QCD phase structure and its relation with the net-baryon number fluctuations in the framework of PNJL quark model. 
The calculation shows that the QCD phase structure is mainly determined by the chiral and deconfinement  transitions, as well as their coincidence and separation. At large chemical potentials, the chiral restoration occurs first with the rise of temperature and then the deconfinement phase takes place at higher temperatures, which forms the quarkyonic phase. At smaller chemical potentials, the deconfinement transition occurs at relatively lower temperatures than that of the chiral  crossover transition. Simultaneously, the  deconfinement transition induces the partial restoration of chiral symmetry, which leads to the chiral crossover transition has a relatively complicated structure as presented in Fig.~2.  For the chiral transition,  there exists a critical end point along  the phase transition line, which connects the  crossover transition at small chemical potentials to the chiral first-order transition at large chemical potentials. 

We plot the contour maps and  the three-dimensional structures of kurtosis and skewness of net baryon number fluctuations. Compared with the experimental data,  our calculations show that the existence of a CEP of chiral transition is essential to explain the large deviation of net-proton kurtosis from the baseline in experiments. Besides, we find that the net-baryon kurtosis and skewness of are closely related to the QCD phase structure determined by the chiral and deconfinement transitions. The measurements of the statistical fluctuation distributions provide a powerful tool to explore the QCD phase structure, including the criticality, the chiral and deconfinement transitions. Further researches on this aspect are needed in both experiments and theories to study the phase structure of strongly interacting matter.
\begin{acknowledgments}
The authors would like to thank Peng-fei Zhuang and Xiao-feng Luo for the useful discussions. This work is supported by the National Natural Science Foundation of China under
Grant No. 11305121, the Natural Science Basic Research Plan
in Shaanxi Province of China (Program No. 2014JQ1012) and the Fundamental Research
Funds for the Central Universities.
\end{acknowledgments}%


\begin{thebibliography}{99}

\bibitem{Gupta11} S. Gupta, X. F. Luo, B. Mohanty, H. G. Ritter, and N. Xu, Science {\bf 332}, 1525  (2011).
\bibitem{Braun09} P. Braun-Munzinger and J. Wambach, Rev. Mod. Phys. {\bf 81}, 1031 (2009).
\bibitem{Fukushima11} K. Fukushima and T. Hatsuda, Rept. Prog. Phys. {\bf 74,} 014001 (2011).
\bibitem{Karsch01} F. Karsch, E. Laermann, and A. Peikert, Nucl. Phys. {\bf B605}, 579 (2001).
\bibitem{Karsch02} F. Karsch, Nucl. Phys.  {\bf A698}, 199 (2002).
\bibitem{Kaczmarek05}O. Kaczmarek and F. Zantow, Phys. Rev. D {\bf 71}, 114510 (2005).
\bibitem{Allton02}C. R. Allton, S. Ejiri, S. J. Hands, O. Kaczmarek, F. Karsch, E. Laermann, C. Schmidt,
and L. Scorzato, Phys. Rev. D {\bf 66}, 074507 (2002).
\bibitem{Cheng06} M. Cheng, N. H. Christ, S. Datta, {\it et al}., Phys. Rev. D {\bf 74}, 054507 (2006).
\bibitem{Aoki09} Y. Aoki, S. Bors\'anyi, S. D\"urr, Z. Fodor, S. D. Katz, S. Krieg, and K. Szabo, J. High Energy Phys. {\bf 06}, 088 (2009).
\bibitem{Fodor07} Z. Fodor, S. D. Katz, and C. Schmidt, J. High Energy Phys. {\bf 03}, 121 (2007) .
\bibitem{Elia09} M. D'Elia and F. Sanfilippo, Phys. Rev. D {\bf 80}, 014502 (2009).
\bibitem{Ejiri08} S. Ejiri, Phys. Rev. D {\bf 78}, 074507 (2008).
\bibitem{Clark07} M. A. Clark and A. D. Kennedy, Phys. Rev. Lett. {\bf 98}, 051601 (2007).
\bibitem{Roberts00}C. D. Roberts and S. M. Schmidt, Prog. Part. Nucl. Phys. {\bf 45}, S1 (2000).
\bibitem{Alkofer01}R. Alkofer and L. von Smekal, Phys. Rep. {\bf 353}, 281 (2001).
\bibitem{Cloet14}I. C. Cl\"{o}t and C. D. Roberts, Prog. Part. Nucl. Phys. {\bf 77}, 1 (2014).
\bibitem{Fischer14}C. S Fischer, J. Luecker, and C. A
Welzbacher. Phys. Rev. D, {\bf 90}, 034022 (2014).
\bibitem{Gao16} F. Gao and Y. X. Liu, Phys.Rev. D {\bf 94}, 076009 (2016). 
\bibitem{Buballa05} M. Buballa, Phys. Rep. {\bf 407}, 205 (2005).
\bibitem{Rehberg96} P. Rehberg, S. P. Klevansky, and J. H\"ufner, Phys. Rev. C {\bf 53}, 410 (1996).
\bibitem{Alford08}  M. Alford, A. Schmit, K. Rajagopal, and T. Sch\"afer, Rev. Mod. Phys. {\bf 80}, 1455 (2008).
\bibitem{Menezes14}D. P. Menezes, M. B. Pinto, L. B. Castro, P. Costa, and C. Provid\^encia, Phys. Rev. C {\bf 89}, 055207  (2014).
\bibitem{Huang16} L. Liu, H. Liu, M. Huang, Phys.Rev. D {\bf 94}, 014026 (2016). 

\bibitem{Fukushima04}  K. Fukushima, Phys. Lett. B {\bf 591}, 277 (2004); Phys. Rev. D 77, 114028 (2008).
\bibitem{Ratti06} C. Ratti, M. A. Thaler, and W. Weise, Phys. Rev. D {\bf 73}, 014019 (2006).
\bibitem{Costa10}  P. Costa, M. C. Ruivo, C. A. de Sousa, and H. Hansen, Symmetry {\bf 2}, 1338 (2010).
\bibitem{Fu08} W. J. Fu, Z. Zhang, and Y. X. Liu, Phys. Rev. D {\bf 77}, 014006 (2008).
\bibitem{Sasaki12} T. Sasaki,J. Takahashi, Y. Sakai,  H. Kouno, and M. Yahiro, Phys. Rev. D {\bf 85}, 056009 (2012).
\bibitem{Ferreira14} M. Ferreira, P. Costa, and C. Provid\^encia, Phys. Rev. D {\bf 89}, 036006 (2014).

\bibitem{Schaefer10} B. J. Schaefer, M. Wagner, and J. Wambach, Phys. Rev. D {\bf 81}, 074013 (2010).
\bibitem{Skokov11}  V. Skokov, B. Friman, and K. Redlich, Phys. Rev. C {\bf 83}, 054904 (2011).
\bibitem{Chatterjee12}  S. Chatterjee and K. A. Mohan, Phys. Rev. D {\bf 85}, 074018, (2012).

\bibitem{Shao11-2} G. Y. Shao, M. Di Toro, B. Liu, M. Colonna, V. Greco, Y. X. Liu, and S. Plumari, Phys. Rev. D {\bf 83}, 094033 (2011).
\bibitem{Shao2015}G. Y. Shao, Z. D. Tang, M. Di Toro, {\it et al.}, Phys. Rev. D {\bf 92}, 114027 (2015).
\bibitem{Shao2016}G. Y. Shao, Z. D. Tang, M. Di Toro, M. Colonna,  X. Y. Gao, N. Gao, Phys. Rev. D {\bf 94}, 014008 (2016).

\bibitem{Aggarwal10} M. M. Aggarwal, {\it et al.}, STAR Collaboration, Phys. Rev. Lett.  {\bf 105}, 022302 (2010) .
\bibitem{Adamczyk14}L. Adamczyk, { \it et al.}, STAR Collaboration, Phys. Rev. Lett. {\bf 112}, 032302 (2014).
\bibitem{Abelev13}B. Abelev, {\it et al.}, ALICE Collaboration, Phys. Rev. Lett. {\bf 110}, 152301 (2013).
\bibitem{Adamczyk142}L. Adamczyk, {\it et al.}, STAR Collaboration, Phys. Rev. Lett.  {\bf 113} 092301 (2014).

\bibitem{Luo2014} X. Luo (for the STAR Collaboration),
PoS(CPOD2014) 019, 2015.
\bibitem{Luo2016} X. Luo, Nucl. Phys. A {\bf 956}, 75 (2016). 
\bibitem{Luo2017} X. Luo and N. Xu, arXiv:1701.02105v1.


\bibitem{Stephanov09} M. A. Stephanov, Phys. Rev. Lett. {\bf 102} 032301 (2009) .
\bibitem{Fu10} W. J. Fu and Y. L. Wu, Phys. Rev. D {\bf 82}, 074013, (2010).
\bibitem{Stephanov11} M. A. Stephanov, Phys. Rev. Lett.  {\bf 107}, 052301 (2011).
\bibitem{Schaefer12}B. J. Schaefer and M. Wagner, Phys. Rev. D  {\bf 85}, 034027 (2012).

\bibitem{Friman11} B. Friman, F. Karsch, K. Redlich, V. Skokov, Eur. Phys.
J. C {\bf 71}, 1694 (2011).
\bibitem{Friman14} B. Friman, Nucl. Phys. A {\bf 928}, 198 (2014). 
\bibitem{Swagato15} S. Mukherjee, R. Venugopalan, and Y. Yin,  Phys. Rev. C {\bf 92}, 034912 (2015).
\bibitem{Chen17}J. W. Chen, J. Deng, H. kohyama, and L. Labun, Phys. Rev. D  {\bf 93}, 034037 (2016); Phys. Rev. D  {\bf 95}, 014038 (2017).
\bibitem{Redlich17} G. A. Alm\'asi, B. Friman, and K. Redlich, arXiv:1703. 0594v1.
\bibitem{Huang17} Z. B. Li, Y. D. Chen, D. N Li, and M. Huang, arXiv:1706.02238v1.
\bibitem{Fan16} W. K. Fan, X. F. Luo, H. S. Zong, arXiv:1608.07903.

\bibitem{Karsch15} H. T. Ding, F. Karsch, and S. Mukherjee. Int. J. Mod. Phys. E {\bf 24}, 530007 (2015).


\bibitem{Robner07} S. R\"{o}{\ss}ner, C. Ratti, and W. Weise, Phys. Rev. D {\bf 75}, 034007 (2007).
\bibitem{Fukugita90} M. Fukugita, M. Okawa, and A. Ukava, Nucl. Phys. B {\bf 337}, 181 (1990).
\bibitem{Cleymans05}J. Cleymans, B. Kampfer, M. Kaneta, S. Wheaton, and N.
Xu, Phys. Rev. C {\bf 71}, 054901 (2005).

\bibitem{Coleman80} S. Coleman and E. Witten, Phys. Rev. Lett. {\bf 45}, 100 (1980).
\bibitem{Fischer142}C. S. Fischer, L. Fister, J. Luecker, and J. M. Pawlowski,
Phys. Lett. B {\bf 732}, 273 (2014).
\bibitem{Xin14} X. Y. Xin, S. X. Qin, and Y. X. Liu, Phys. Rev. D {\bf 89}, 094012 (2014).
\bibitem{Kitazawa12} M. Kitazawa and M. Asakawa, Phys. Rev. C {\bf 85}, 021901 (2012);
 {\bf86}, 024904 (2012);  {\bf 86}, 069902 (2012).
\bibitem{fukushima15} K. Fukushima, Phys. Rev. C {\bf 91}, 044910 (2015)

%
%

\end{thebibliography}
\end{document}